\documentclass[aps,prd,reprint,amsmath,amssymb,nofootinbib]{revtex4-2}
\usepackage[hidelinks]{hyperref}
\newcommand{\R}{\mathcal R}
\begin{document}
\title{Comment on ``Quantum electron clouds near black holes: Black atoms and molecules''}
\author{Khireddine Nouicer}
\email{khnouicer@univ-jijel.dz}
\thanks{ORCID: 0000-0001-6110-9833}
\affiliation{Laboratory of Theoretical Physics and Department of Physics, Faculty of Exact and Computer Sciences, University of Jijel, BP 98 Ouled Aissa, Jijel, 18000, Algeria}
% Optional present/secondary address, if applicable:
% \altaffiliation{Present address: DEPARTMENT, INSTITUTION, CITY, COUNTRY}
% Authenticate the ORCID iD in the APS submission system.
\begin{abstract}
We revisit the curvature-dependent Schr\"odinger equation employed by Iseki, Sasaki, and Shiozawa [Phys. Rev. D \textbf{113}, 025015 (2026)] to describe quantum electron clouds in black-hole geometries. Restoring the missing inverse-mass factor in the curvature coupling and consistently enforcing the positive sign prescription of that work alters the radial equation in the Reissner--Nordstr\"om background. On domains reaching the central singularity, the shifted inverse-square potential crosses the critical Hardy coupling: the overextremal radial operator of that work is unbounded from below and has no ground state, whereas the corrected Hamiltonian admits a semibounded Friedrichs extension. We further identify the continuum threshold associated with the extremal throat, showing that the reported $3p$ energy lies within the throat continuum and thus cannot support a normalizable exterior bound state. Finally, accounting for the radial integration measure clarifies the physical picture of spatial localization.
 In the multicenter Majumdar--Papapetrou geometry, the corrected curvature term survives the formal small-mass limit at fixed background, while the spheroidal separation requires modifications that affect the separated far-field solutions. These findings suggest that the spectral interpretation of atomic and molecular configurations in the original nonrelativistic model should be reassessed.
\end{abstract}
\maketitle
\section{Curvature-dependent quantization}
We examine how inconsistencies in the spatial Schr\"odinger model of Ref.~\cite{ISS} affect spectral stability and the existence of normalizable bound states, retaining its classical backgrounds and potentials. All equation and figure numbers cited from that work refer to the published version, Phys. Rev. D \textbf{113}, 025015 (2026), published on 26 January 2026.

The Lagrangian (2.1) of Ref.~\cite{ISS} contains a particle mass $\mu$ multiplying the spatial metric $g_{ij}$. In Eqs.~(2.4), (2.10), and (2.11), the curvature correction lacks the corresponding inverse mass. With the positive prescription chosen in footnote 1 of Ref.~\cite{ISS}, the dimensionally consistent Hamiltonian is instead
\begin{equation}\label{H}
 H_+=-\frac{\hbar^2}{2\mu}\left(\Delta_g-\frac{R^{(3)}}6\right)+V,
\end{equation}
where $g=\det(g_{ij})$, $\Delta_g=g^{-1/2}\partial_i(\sqrt g\,g^{ij}\partial_j)$ and $R^{(3)}$ is the spatial scalar curvature. The factor $1/\mu$ converts $\hbar^2R^{(3)}$ to energy units.
This normalization also follows by absorbing the mass into the configuration metric $G_{ij}=\mu g_{ij}$. The connection is unchanged under this constant rescaling, whereas
\begin{equation}
 \Delta_G=\mu^{-1}\Delta_g,\qquad R[G]=\mu^{-1}R[g].
\end{equation}
The numerical coefficient is prescription-dependent, as discussed in Refs.~\cite{DeWitt,Parker,Assirati}; in particular, the positive DeWitt coefficient with the inverse mass appears in Eq.~(11) of Ref.~\cite{Assirati}. Our criticism concerns consistency with the prescription selected in Ref.~\cite{ISS}, not the uniqueness of that prescription. The two signs need not be spectrally equivalent. In what follows we use $H_\zeta=-\hbar^2(\Delta_g-\zeta R^{(3)})/(2\mu)+V$ to compare them.

\section{Radial measure and boundary conditions}
For a spatial metric $dl^2 = dr^2/f + r^2\,d\Omega^2$, expanding in normalized spherical harmonics yields the radial integration measure $r^2\,dr/\sqrt{f}$. The proper radial distance $s$ and the reduced radial wave function $v$ satisfy
\begin{equation}\label{unitary}
 ds=\frac{dr}{\sqrt f},\quad v=r\R_\ell,\quad
 \int\frac{r^2}{\sqrt f}|\R_\ell|^2dr=\int|v|^2ds.
\end{equation}
Since $dr/ds=\sqrt f$ and $d^2r/ds^2=f'/2$, differentiation of $\R_\ell=v/r$ yields
\begin{equation}
 \left(\partial_s^2+\frac{2\dot r}{r}\partial_s\right)\frac vr
 =\frac1r\left(v''-\frac{f'}{2r}v\right),
\end{equation}
where the dot and the derivatives of $v$ refer to $s$, and $f'=df/dr$. At a simple exterior horizon $r_h$, $s-s_h\sim 2\sqrt{(r-r_h)/f'(r_h)}$. Thus the constant and square-root branches discussed after Eq.~(3.6) and in Eq.~(4.7) of Ref.~\cite{ISS} are both regular in proper distance. More explicitly, if $\R_\ell=A+B\sqrt{r-r_h}+O(r-r_h)$, then
\begin{equation}
 \R_\ell=A+\frac{B\sqrt{f'(r_h)}}2(s-s_h)+O((s-s_h)^2).
\end{equation}
The apparent divergence of the coordinate derivative on the second branch does not exclude it. A well-posed self-adjoint radial problem requires specifying a boundary condition at this regular endpoint, such as a Robin condition $v'(s_h) = \lambda v(s_h)$ with $\lambda \in \mathbb{R}$, or Dirichlet data. These represent self-adjoint extensions of the spatial operator rather than a prescription for absorption across a Lorentzian horizon.

In Schwarzschild, $f=1-a/r$ and $R^{(3)}=0$. A nonzero finite horizon value of $\R_\ell$ produces a divergent coordinate density $r^2|\R_\ell|^2/\sqrt f$, but the proper-distance density $|v|^2$ remains finite. Although Ref.~\cite{ISS} identifies the metric origin of the divergence, its interpretation as trapping requires additional evidence. For a bounded radial function the integrated probability satisfies
\begin{equation}
 \int_{r_h}^{r_h+\delta}\frac{r^2}{\sqrt f}|\R_\ell|^2dr=O(\sqrt\delta)\longrightarrow0.
\end{equation}
Consequently the peak alone is insufficient to establish trapping. The free positive-energy solutions are scattering states. Dividing Eq.~(3.6) of Ref.~\cite{ISS} by $f$ and setting $\R=f^{-1/4}Z$ removes the extra first derivative, giving
\begin{equation}
 Z''+\frac2r Z'+\left[\frac{k^2}{f}-\frac{\ell(\ell+1)}{r^2 f}
 +\frac{3a^2}{16r^4 f^2}\right]Z=0,
\end{equation}
where $k^2=2\mu E/\hbar^2$. This substitution is the inverse of the measure-flattening rescaling $\widehat R=f^{-1/4}\R$ in footnote 2 of Ref.~\cite{ISS}. Expanding the coefficient of $Z$ at fixed $a,k$ gives
\begin{equation}
 k^2+\frac{ak^2}{r}+\frac{a^2k^2-\ell(\ell+1)}{r^2}+O(r^{-3}).
\end{equation}
Equation (3.7) of Ref.~\cite{ISS} omits $a^2k^2/r^2$ despite stating that only $O(r^{-3})$ terms were discarded; a separate truncation linear in $a$ would be needed. The asymptotic interpretation following its Eq.~(3.8) also misses a logarithmic phase. Since the reduced free potential is $O(s^{-2})$ and $s=r+(a/2)\ln(r/r_0)+O(1)$, the two asymptotic branches for $k=\sqrt{2\mu E}/\hbar>0$ are
\begin{equation}
 \R_\pm\sim\frac{C_\pm}{r}\exp\left[\pm i\left(kr+\frac{ak}{2}\ln\frac r{r_0}\right)\right].
\end{equation}
The algebraic Heun branch in Ref.~\cite{ISS} independently gives the power $-(2\sigma+a^2k^2)/(2\sigma)=-1\pm iak/2$, with $\sigma=\pm iak$.
The effective $1/r$ term in the coordinate equation is energy-dependent:
\begin{equation}
 V_{\rm eff}=-\frac{\hbar^2ak^2}{2\mu r}=-\frac{aE}{r},\qquad V_{\rm Newt}=-\frac{GM_{\rm phys}\mu}{r}.
\end{equation}
For $a=2GM_{\rm phys}/c^2$, these differ by the factor $2E/(\mu c^2)$, so the coordinate term should not be identified with the Newtonian interaction. Likewise, asymptotic flatness does not enforce the flat-space hydrogen spectrum imposed before Eqs.~(3.10) and (4.8) of Ref.~\cite{ISS}: decay at infinity must be combined with the boundary condition at the other endpoint to determine the eigenvalues. The exponential substitution printed immediately before Eq.~(3.10) also contains an extraneous imaginary unit: the displayed Heun equation follows from $\R(x)=e^{\sigma x}H(x)$ with $\sigma=\pm a\kappa$ and $\kappa=\sqrt{-2\mu E}/\hbar$, not from $e^{i\sigma x}H(x)$. Equation (3.10) itself is consistent with the corrected substitution.

\section{Reissner--Nordstr\"om geometry}
For $f=1-a/r+Q^2/r^2$ and $V=-e^2/r$ (the conventions of Ref.~\cite{ISS}), the scalar curvature is $2(1-f-rf')/r^2=2Q^2/r^4$, in agreement with its Eq.~(4.3). The stationary equation obtained from Eq.~\eqref{H} has a positive potential $\hbar^2Q^2/(6\mu r^4)$, correcting Eq.~(4.4). More generally, separating $H_\zeta$ gives
\begin{align}\label{rad}
 0={}&f\R_\ell''+\left(\frac{2f}{r}+\frac{f'}2\right)\R_\ell'\nonumber\\
 &+\left[\frac{2\mu}{\hbar^2}\left(E+\frac{e^2}{r}\right)
 -\frac{\ell(\ell+1)}{r^2}-\frac{2\zeta Q^2}{r^4}\right]\R_\ell .
\end{align}
Here derivatives are in $r$. The positive prescription fixes $\zeta=1/6$, whereas Eq.~(4.5) of Ref.~\cite{ISS} corresponds to $\zeta=-1/6$. In Eq.~(4.8) of Ref.~\cite{ISS}, the corresponding replacement is
\begin{equation}
 \frac{4\kappa^2Q^2}{\rho^2}(\ell^2+1/3)\ \longrightarrow\ \frac{4\kappa^2Q^2}{\rho^2}(\ell^2-1/3),
\end{equation}
with $\kappa=\sqrt{-2\mu E}/\hbar$ and $\rho=2\kappa r$. The Coulomb parameter $\mu e^2/(\hbar^2\kappa)$ need not be an integer. These changes concern the equation given in Ref.~\cite{ISS} in connection with Fig.~6; their numerical impact depends on the actual implementation.

For $Q\ne0$, the leading coefficients near the central singularity are $f\sim Q^2/r^2$ and $2f/r+f'/2\sim Q^2/r^3$. Multiplying Eq.~\eqref{rad} by $r^4/Q^2$ gives the dominant equation and its indicial condition,
\begin{equation}
 r^2\R''+r\R'-2\zeta\R=0,\qquad \sigma^2=2\zeta.
\end{equation} The corrected powers are $r^{\pm1/\sqrt3}$, whereas the opposite sign yields $\exp[\pm i\ln(r/r_0)/\sqrt3]$. In particular, the three indicial results are
\begin{equation}
 \begin{array}{ll}
 i/\sqrt{3Q} &\text{Eq.~(4.6) of Ref.~\cite{ISS}},\\
 \pm i/\sqrt3 &\text{derived from Eq.~(4.5) of Ref.~\cite{ISS}},\\
 \pm 1/\sqrt3 &\text{corrected equation with }\zeta=1/6.
 \end{array}
\end{equation}
The charge dependence of the first expression does not follow from the radial equation and is dimensionally inconsistent for a geometrical charge, and the bounded oscillatory branches do not have a limit at the origin.

\subsection{Spectral stability}
For the overextremal case ($Q^2>a^2/4$), the unitary map \eqref{unitary} takes Eq.~\eqref{rad} to $-v''+W_\zeta v=(2\mu E/\hbar^2)v$, where
\begin{align}\label{W}
 W_\zeta={}&\frac{\ell(\ell+1)}{r^2}+\frac{a}{2r^3}
 +(2\zeta-1)\frac{Q^2}{r^4}-\frac{2\mu e^2}{\hbar^2r}.
\end{align}
Near the central endpoint, expansion of the proper distance gives
\begin{equation*}
 s=\frac{r^2}{2|Q|}+O(r^3),\quad
 \frac{Q^2}{r^4}=\frac1{4s^2}+O(s^{-3/2}).
\end{equation*}
The other contributions in Eq.~\eqref{W} are less singular; hence
\begin{equation}
 W_\zeta=\frac{\gamma_\zeta}{s^2}+O(s^{-3/2}),\qquad
 \gamma_\zeta=\frac{2\zeta-1}{4}.
\end{equation}
The two signs therefore lie on opposite sides of the critical coupling:
\begin{equation}\label{hardy}
 \gamma_-=-\frac13<-\frac14<-\frac16=\gamma_+.
\end{equation}
To establish the spectral consequence, define the reduced quadratic form on smooth compactly supported functions by
\begin{equation}
 \mathfrak q_\zeta[v]=\int (|v'|^2+W_\zeta|v|^2)ds.
\end{equation}
The physical energy form in a fixed angular channel is $\hbar^2\mathfrak q_\zeta/(2\mu)$. The optimal Hardy inequality \cite{Hardy} is
\begin{equation}
 \int_0^\infty|v'|^2ds\ge\frac14\int_0^\infty\frac{|v|^2}{s^2}ds.
\end{equation}
For the corrected sign the leading form thus controls $\tfrac13\int|v'|^2ds$. Its subleading potential is bounded locally in absolute value by $\delta/s^2+C_\delta$ for arbitrary $\delta>0$, and can be absorbed into this estimate. A global lower bound follows directly from the non-negativity of the kinetic contribution for $\zeta=1/6$. The potential energy reads
\begin{equation*}
 \begin{gathered}
 V_+(r)=\frac{A}{r^4}-\frac{B}{r},\\
 A=\frac{\hbar^2Q^2}{6\mu},\qquad B=e^2>0.
 \end{gathered}
\end{equation*}
The stationarity condition,
\begin{equation*}
 V_+'(r_*)=0\quad\Longrightarrow\quad r_*=\left(\frac{4A}{B}\right)^{1/3},
\end{equation*}
determines the unique global minimum of the potential,
\begin{equation*}
 V_+(r_*)=-\frac{3B}{4}\left(\frac{B}{4A}\right)^{1/3}.
\end{equation*}
The Friedrichs extension, defined by closure of the lower-bounded form \cite{ReedSimon}, therefore satisfies
\begin{equation}\label{lower}
 H_+^F\ge-\frac{3e^2}{4}\left(\frac{3\mu e^2}{2\hbar^2Q^2}\right)^{1/3}.
\end{equation}

For $\zeta=-1/6$, optimality of Hardy's constant ensures a normalized $\phi\in C_c^\infty(0,\infty)$ for which $\int(|\phi'|^2-|\phi|^2/(3s^2))ds=-C<0$. The normalized functions $\phi_\lambda(s)=\sqrt\lambda\,\phi(\lambda s)$ concentrate near the center. Under $x=\lambda s$, both the derivative and inverse-square integrals scale as $\lambda^2$. Writing $W_-=-1/(3s^2)+\mathcal W$, with $\mathcal W=O(s^{-3/2})$, gives
\begin{equation}
 \int\mathcal W(s)|\phi_\lambda(s)|^2ds
 =\int\mathcal W(x/\lambda)|\phi(x)|^2dx
 =O(\lambda^{3/2}),
\end{equation}
since $\phi$ has compact support away from zero. Hence
\begin{equation}\label{collapse}
 \mathfrak q_-[\phi_\lambda]=-C\lambda^2+O(\lambda^{3/2})
 \longrightarrow-\infty.
\end{equation}
Since the normalized test functions $\phi_\lambda$ belong to the domain of every self-adjoint extension of the minimal radial operator, Eq.~\eqref{collapse} implies
\begin{equation}
 \inf\sigma(H_-^{\rm SA})=-\infty,
\end{equation}
where $H_-^{\rm SA}$ denotes an arbitrary self-adjoint realization of the radial Hamiltonian with $\zeta=-1/6$ on an interval reaching the central singularity. In particular, the overextremal configuration in Sec.~IV C of Ref.~\cite{ISS} cannot have a ground state with this curvature coupling, although individual stationary solutions may exist. This statement concerns the spatial nonrelativistic model extrapolated to the center, not the exterior of a subextremal black hole or the full relativistic theory. In particular, the central region need not satisfy the local nonrelativistic and tidal conditions stated in the introduction of Ref.~\cite{ISS}; the result is a statement about the extrapolated operator, not a prediction of physical collapse in the relativistic regime.

The boundary conditions do not remove this distinction. Indeed, the local indicial equation for $v$ is $\beta(\beta-1)=\gamma_\zeta$ (equivalently one can use Frobenius in $\sqrt s$), yielding
\begin{equation}
 \begin{aligned} v_\pm&\sim s^{1/2\pm\alpha}\quad(\zeta=1/6),\\
 v_\pm&\sim s^{1/2\pm i\alpha}\quad(\zeta=-1/6),\end{aligned}
\end{equation}
where $\alpha=1/(2\sqrt3)$ and a reference length is implicit in the powers. Both branches are locally square integrable for either sign, so the central endpoint is limit-circle. In the overextremal problem infinity is limit-point and the radial deficiency indices are $(1,1)$ \cite{Case,Frank,ReedSimon}. For the corrected sign the Friedrichs condition eliminates the $s^{1/2-\alpha}$ branch: the leading derivatives satisfy
\begin{equation}
 |v_-'|^2\sim C_-s^{-1-2\alpha},\qquad |v_+'|^2\sim C_+s^{-1+2\alpha}.
\end{equation}
Only the second expression is integrable. The local Hardy estimate, after absorbing the subleading potential, controls the local $H^1$ norm after adding a sufficiently large multiple of the squared $L^2$ norm to the closed form. The principal branch is admissible there by approximation with smooth cutoffs. This agrees with the principal-solution characterization of the Friedrichs extension \cite{GLN}. In the supercritical case, choosing a self-adjoint boundary condition between the oscillatory branches cannot alter Eq.~\eqref{collapse}.

\subsection{Exactly extremal geometry}
The simple-horizon roots $0,1/2$ in Eq.~(4.7) of Ref.~\cite{ISS} cannot be carried over to its Sec.~IV B. Writing $M=GM_{\rm phys}/c^2$ for the geometrical mass, at $a=2M$, $Q^2=M^2$, one has $f=(1-M/r)^2$. For $x=r-M>0$, $f\sim x^2/M^2$ and $2f/r+f'/2\sim x/M^2$. Substitution of $\R\sim x^\sigma$ into Eq.~\eqref{rad} then gives
\begin{align}
 \sigma^2&=\frac{2\mu M^2}{\hbar^2}(E_H-E),\\
 E_H&=\frac{\hbar^2}{2\mu M^2}[\ell(\ell+1)+2\zeta]-\frac{e^2}{M}.
\end{align}
The exterior proper coordinate $s=r+M\ln[(r-M)/M]+\mathrm{const}$ ranges over the real line. In energy units the potential $\hbar^2 W_\zeta/(2\mu)$ tends to $E_H$ at the throat and to zero at spatial infinity. Both ends are limit-point. Comparing with a step potential with these two limits gives the essential spectrum of the radial Hamiltonian, by Weyl's theorem \cite{RS4} (Theorem XIII.14):
\begin{equation}\label{ess}
 \sigma_{\rm ess}(H_{\ell,\zeta})=[\min(0,E_H),\infty).
\end{equation}
Here, $\sigma_{\rm ess}$ denotes the essential spectrum, which excludes isolated eigenvalues of finite multiplicity. Indeed, decoupling the step operator at a finite point changes its resolvent by a finite-rank operator, and the difference between the full and step potentials is bounded and vanishes at both ends, rendering it relatively compact. Any discrete eigenvalues must therefore lie strictly below both thresholds. The sign correction shifts the throat threshold by
\begin{equation}
 E_H^{(+)}-E_H^{(-)}=\frac{\hbar^2}{3\mu M^2},
\end{equation}
so an energy evanescent in the corrected throat may lie in the propagation continuum for $\zeta=-1/6$.

For example, writing $g_C=2\mu Me^2/\hbar^2$, the flat hydrogen energies used in Ref.~\cite{ISS} satisfy $2\mu M^2 E_n/\hbar^2=-g_C^2/(4n^2)$. A necessary condition for such an energy to lie below the corrected throat continuum is
\begin{equation}
 -\frac{g_C^2}{4n^2}<\ell(\ell+1)+\frac13-g_C.
\end{equation}
Even when this holds, the global eigenvalue condition must still be imposed. In particular, Figs.~8 and 9 of Ref.~\cite{ISS} use $a=6$, $Q=3$ at extremality and the units $\mu=e=\hbar=1$ specified in Fig.~3. For $M=3$, the energies and throat thresholds in the $3p$ and $5d$ channels of Fig.~8 are
\begin{equation}
\begin{array}{c|ccc}
 (n,\ell)&E_n&E_H^{(-)}&E_H^{(+)}\\ \hline
 (3,1)&-1/18&-13/54&-11/54\\
 (5,2)&-1/50&-1/54&1/54
\end{array}
\end{equation}
The $3p$ energy lies strictly above both throat thresholds and therefore cannot support a square-integrable exterior mode, as the local analysis below establishes. The $5d$ energy lies below both continuum thresholds, but this alone does not establish an eigenvalue.

For an individual Frobenius branch $\R\sim Cx^\sigma$ ($C\ne0$), the radial norm near the extremal horizon behaves as
\begin{equation}
 \int_0^\delta \frac{(M+x)^3}{x}|\R(M+x)|^2dx
 \ \sim\ \mathrm{const}\int_0^\delta x^{2\operatorname{Re}\sigma-1}dx.
\end{equation}
For $E<E_H$, only the positive-root branch is square-integrable. At $E=E_H$, the independent constant and logarithmic leading behaviors both give divergent norms. For $E>E_H$, writing $t=-\ln(x/M)$, a general nonzero solution has leading behavior $A e^{i\omega t}+B e^{-i\omega t}$, $\omega=\sqrt{2\mu M^2(E-E_H)}/\hbar>0$, with exponentially decaying Frobenius remainders. Its squared modulus has positive period average $|A|^2+|B|^2$; the norm therefore diverges also for every nontrivial linear combination. Consequently, a wave function that approaches a finite, nonzero value at the extremal horizon cannot represent a normalizable bound state in this measure. This distinction is essential when interpreting the profiles in Figs.~8 and 9 of Ref.~\cite{ISS}.

\section{Multicenter geometries}
For the Majumdar--Papapetrou spatial metric $g_{ij}=U^2\delta_{ij}$, with $U=1+\sum_a M_a/|\mathbf x-\mathbf x_a|$, direct differentiation gives, away from the centers,
\begin{align}
 \Delta_g\psi&=U^{-3}\nabla\cdot(U\nabla\psi),\\
 R^{(3)}&=-4U^{-3}\Delta U+2U^{-4}|\nabla U|^2
 =2U^{-4}|\nabla U|^2.
\end{align}
In the printed Eq.~(5.4) of Ref.~\cite{ISS}, the curvature contribution lacks both the inverse mass and the multiplicative wave function. The homogeneous eigenvalue equation requires the term $\hbar^2 R^{(3)}\psi/(12\mu)$. Setting $\psi=U^{-1/2}\chi$ and applying the product rule to $\Delta_g=U^{-2}\Delta+U^{-3}\nabla U\cdot\nabla$, with $\Delta U=0$, cancels the first derivatives of $\chi$ and gives
\begin{equation}
 \Delta_g\psi=U^{-5/2}\left[\Delta\chi+\frac14|\nabla\ln U|^2\chi\right].
\end{equation}
Combining this identity with the curvature term in Eq.~\eqref{H}, the corrected Eq.~(5.4) of Ref.~\cite{ISS} becomes
\begin{equation}\label{mp}
 \left[-\Delta+\frac1{12}|\nabla\ln U|^2\right]\chi
 =\frac{2\mu U^2}{\hbar^2}(E-V)\chi.
\end{equation}
On $\Sigma=\mathbb R^3\setminus\{\mathbf x_1,\ldots,\mathbf x_N\}$ the transformation preserves the norm,
\begin{equation}
 \|\psi\|_g^2=\int_\Sigma U^3|\psi|^2d^3x=\int_\Sigma U^2|\chi|^2d^3x,
\end{equation}
so the transformed Hilbert space is $\mathcal H=L^2(\Sigma,U^2d^3x)$. In the formal limit $\mu(E-V)\to0$ locally at fixed geometry, the curvature contribution does not vanish. The limiting operator, initially defined on the dense domain $C_c^\infty(\Sigma)$ in the weighted Hilbert space $\mathcal H$, is
\begin{equation}
 \mathcal A_0=U^{-2}[-\Delta+\tfrac1{12}|\nabla\ln U|^2].
\end{equation}
Its Friedrichs form is the closure of the nonnegative expression
\begin{equation}
 \int\left(|\nabla\chi|^2+\frac1{12}|\nabla\ln U|^2|\chi|^2\right)d^3x.
\end{equation}
A zero mode in its operator domain would have zero form and hence be constant, which is incompatible with the norm at infinity unless it vanishes. Thus, the Friedrichs realization $\mathcal A_0^F$ has a trivial kernel, $\ker\mathcal A_0^F=\{0\}$, and therefore admits no nontrivial normalizable zero mode.

For two equal centers separated by $d$, let $D=\xi^2-\eta^2$, $b=4M/d$, and $U=1+b\xi/D$ in the prolate spheroidal coordinates of Ref.~\cite{ISS}. The derivatives are $U_\xi=-b(\xi^2+\eta^2)/D^2$ and $U_\eta=2b\xi\eta/D^2$. The Euclidean scale factors are
\begin{equation}
 \begin{aligned}
 h_\xi^2&=\frac{d^2D}{4(\xi^2-1)},& h_\eta^2&=\frac{d^2D}{4(1-\eta^2)},\\
 h_\phi^2&=\frac{d^2}{4}(\xi^2-1)(1-\eta^2).
 \end{aligned}
\end{equation}
Since $U$ is independent of the azimuthal coordinate $\phi$, the contribution $h_\phi^{-2}(\partial_\phi U)(\partial_\phi\psi)$ vanishes identically, including for $m\ne0$. Using the inverse squared scale factors then yields
\begin{align}
 \nabla U\cdot\nabla\psi={}&\frac4{d^2D}[(\xi^2-1)U_\xi\psi_\xi\nonumber\\
 &+(1-\eta^2)U_\eta\psi_\eta],\\
 R^{(3)}={}&\frac{128M^2(\xi^4+3\xi^2\eta^2-\xi^2+\eta^2)}{(4M\xi+dD)^4}.
\end{align}
These expressions correct the missing factor four in Eq.~(5.10) and the denominator sign in Eq.~(5.8) of Ref.~\cite{ISS}. The corrected denominator is proportional to $U^4$ after restoring the common factor $(dD)^4$, and has no zero away from the centers for positive $M,d$. Writing the wave function as $\psi(\xi,\eta,\phi)=e^{im\phi}\psi_m(\xi,\eta)$, with $m\in\mathbb Z$, the Euclidean Laplacian in prolate spheroidal coordinates reduces to
\begin{equation*}
 \Delta\psi=\frac{4e^{im\phi}}{d^2D}\mathcal P_m\psi_m,
\end{equation*}
where
\begin{align}
 \mathcal P_m={}&\partial_\xi[(\xi^2-1)\partial_\xi]+
 \partial_\eta[(1-\eta^2)\partial_\eta]\nonumber\\
 &-m^2[(\xi^2-1)^{-1}+(1-\eta^2)^{-1}].
\end{align}
Multiplying the corrected stationary equation for $\psi_m$ by $U^2d^2D/4$ gives the replacement for Eq.~(5.9) of Ref.~\cite{ISS}:
\begin{align}
 0={}&\mathcal P_m\psi_m+\frac{(\xi^2-1)U_\xi\partial_\xi\psi_m+(1-\eta^2)U_\eta\partial_\eta\psi_m}{U}\nonumber\\
 &+\frac{\mu d^2D U^2}{2\hbar^2}(E-V)\psi_m\nonumber\\
 &-\frac{(\xi^2-1)U_\xi^2+(1-\eta^2)U_\eta^2}{3U^2}\psi_m .
\end{align}
Relative to Eq.~(5.9) of Ref.~\cite{ISS}, the first-derivative coefficients acquire a factor four and the angular derivative changes sign; the energy coefficient acquires a factor $1/4$, and the curvature term has corrected normalization and denominator. We adopt the stipulated Coulomb potential $V=-e^2/R_1-e^2/R_2=-4e^2\xi/(dD)$, consistently with the definition of $A$ in Ref.~\cite{ISS}. With this convention, the energy and Coulomb terms combine as $U^2(dA\xi-\kappa^2d^2D/4)$ with $A=2\mu e^2/\hbar^2$ and $E<0$.

If one subsequently makes the far-field coefficient truncation of Sec.~V of Ref.~\cite{ISS}, the radial equation replacing Eq.~(5.11) reads
\begin{equation}\label{trunc}
 [(\xi^2-1)F']'-bF'-\left(\Lambda+\frac{m^2}{\xi^2-1}\right)F=0.
\end{equation}
In particular, the radial derivative coefficient is $-4M/d$, not $-M/d$. Setting $F = hY$ and eliminating the first-derivative term yields $h'/h = b/[2(\xi^2 - 1)]$.
Integrating for $h$ and substituting into the differential equation yields
\begin{align}
 h&=\left(\frac{\xi-1}{\xi+1}\right)^{b/4},\\
 0&=(\xi^2-1)Y''+2\xi Y'-\left(\Lambda+\frac{m^2+b^2/4}{\xi^2-1}\right)Y.
\end{align}
Thus, the corrected counterpart of Eq.~(5.12) for the truncated equation is
\begin{align}
 F(\xi)&=\left(\frac{\xi-1}{\xi+1}\right)^{M/d}
 [C_1P_\nu^\beta(\xi)+C_2Q_\nu^\beta(\xi)],\\
 \nu(\nu+1)&=\Lambda,\qquad \beta^2=m^2+4M^2/d^2.
\end{align}
Here, $P_\nu^\beta(\xi)$ and $Q_\nu^\beta(\xi)$ denote the associated Legendre functions of the first and second kinds, respectively, with degree $\nu$ and order $\beta$.

For exceptional parameter values at which the two associated Legendre functions become linearly dependent, a second linearly independent solution may be obtained by an appropriate limiting procedure. The discrepancy with Eq.~(5.12) of Ref.~\cite{ISS} persists even if one retains the coefficient $b_{\rm ISS}=M/d$. In that case, the radial equation requires a prefactor exponent $M/(4d)$ and a squared Legendre order $\beta^2=m^2+M^2/(16d^2)$, rather than the value $m^2+M^2/(4d^2)$ reported in Ref.~\cite{ISS}. Although regularity of the angular solution imposes $\Lambda=L(L+1)$, with integer $L\geq |m|$, this condition does not imply that the order $\beta$ of the radial Legendre functions must also be an integer.

Importantly, Eq.~\eqref{trunc} is only a formal truncation. At fixed $b$, the far-field expansions are
\begin{align}
 U&=1+\frac b\xi+\frac{b\eta^2}{\xi^3}+O(\xi^{-5}),\\
 \frac{(\xi^2-1)U_\xi}{U}&=-b+\frac{b^2}{\xi}+O(\xi^{-2}),\\
 C_R&=-\frac{b^2}{3\xi^2}+O(\xi^{-3}),
\end{align}
where $C_R$ denotes the multiplicative curvature coefficient in the equation for $\psi_m$. For a power-law branch the omitted terms contribute at the same order as the retained $m^2/(\xi^2-1)$ term. The separated Legendre solutions therefore do not provide a systematically complete expansion at that order, nor establish a molecular spectrum. At finite mass the energy term also grows quadratically in $\xi$, so the far-field and small-mass limits need not commute.

\begin{table*}[t]
\caption{Explicit corrections to the published equations of Ref.~\cite{ISS}. The final column identifies the corresponding equations in this Comment. Here $D=\xi^2-\eta^2$, $b=4M/d$, and the positive curvature prescription is retained. The last two rows concern only the formal far-field truncation.}\label{tab:corrections}
\begingroup\small\setlength{\tabcolsep}{5pt}\renewcommand{\arraystretch}{1.27}
\begin{tabular}{p{0.13\textwidth}p{0.29\textwidth}p{0.42\textwidth}p{0.08\textwidth}}
\hline\hline
Ref.~[1] & Inconsistency & Correction & This work \\\hline
(2.4), (2.10), (2.11) & Missing inverse-mass factor in curvature energy & $\hbar^2 R^{(3)}/12\ \to\ \hbar^2 R^{(3)}/(12\mu)$ & (1), (2) \\
(3.7) & Omission at order $r^{-2}$ & Restore $a^2k^2/r^2$ at fixed $a,k$ & (7), (8) \\
After (3.8) & Missing logarithmic scattering phase & $R_\pm\sim (C_\pm/r)\exp\{\pm i[kr+(ak/2)\ln(r/r_0)]\}$ & (9) \\
Before (3.10) & Extraneous imaginary unit in exponential substitution & $R=e^{i\sigma x}H\ \to\ R=e^{\sigma x}H$; Eq.~(3.10) itself is unchanged & Sec.~II \\
(4.4) & Missing inverse-mass factor & $\hbar^2 Q^2/(6r^4)\ \to\ \hbar^2 Q^2/(6\mu r^4)$ & (11) \\
(4.5) & Curvature sign inconsistent with positive prescription & $+Q^2/(3r^4)\ \to\ -Q^2/(3r^4)$ in the radial equation & (11) \\
(4.6) & Incorrect charge dependence of the indicial exponent & $\sigma=\pm i/\sqrt3$ for (4.5); $\sigma=\pm1/\sqrt3$ for the corrected sign & (13), (14) \\
(4.8) & Incorrect curvature sign & $\ell^2+1/3\ \to\ \ell^2-1/3$ in the $4\kappa^2Q^2/\rho^2$ term & (12) \\
(5.4) & Missing inverse-mass factor and multiplicative wave function & $\hbar^2R^{(3)}\psi/(12\mu)$; transformed equation with $|\nabla\ln U|^2/12$ & (33)--(36) \\
(5.8) & Wrong sign in curvature denominator & $(4M\xi-dD)^4\ \to\ (4M\xi+dD)^4$ & (42) \\
(5.9) & Incorrect gradient, energy and curvature coefficients & Gradient terms: factor four and reversal of $\eta$ sign; energy term: $-\kappa^2d^2D/4$; curvature: Eq.~(44) & (44) \\
(5.10) & Missing overall factor four & $1/d^2\ \to\ 4/d^2$; azimuthal contribution vanishes since $U_\phi=0$ & (40), (41) \\
(5.11) & Incorrect radial derivative coefficient & $-M/d\ \to\ -4M/d$ & (45) \\
(5.12) & Radial prefactor and Legendre order & $h=[(\xi-1)/(\xi+1)]^{M/d}$, $\beta^2=m^2+4M^2/d^2$ & (46)--(49) \\
\hline\hline
\end{tabular}\endgroup
\end{table*}

\begin{table*}[t]
\caption{Spectral and physical implications within the spatial nonrelativistic model of Ref.~\cite{ISS}. These statements concern the printed equations and not an independently verified numerical implementation.}\label{tab:physics}
\begingroup\small\setlength{\tabcolsep}{5pt}\renewcommand{\arraystretch}{1.25}
\begin{tabular}{p{0.20\textwidth}p{0.30\textwidth}p{0.42\textwidth}}
\hline\hline
Issue & Interpretation or assumption in Ref.~[1] & Revised conclusion \\\hline
Horizon localization & Divergent radial coordinate density interpreted as trapping & Proper-distance density is finite for bounded simple-horizon solutions; integrated shell probability vanishes as the shell shrinks. \\
Simple-horizon endpoint & Regularity of radial wave function & Both local branches are regular in proper distance; an endpoint condition is needed for a self-adjoint radial problem. \\
Newtonian limit & Effective $1/r$ term identified with gravity & $V_{\rm eff}=-aE/r$ is energy dependent and differs from $-GM_{\rm phys}\mu/r$. \\
Hydrogenic energies & Flat-space values inferred from asymptotic flatness & Both endpoint conditions determine the eigenvalues. \\
Central RN stability & Stationary profiles on domains reaching the center & Printed negative sign gives $\gamma_-=-1/3<-1/4$ (no ground state); corrected positive sign gives $\gamma_+=-1/6>-1/4$ (semibounded Friedrichs extension). \\
Extremal RN & Reported exterior $3p$ profile interpreted as bound & Its assigned energy is in the throat continuum for both curvature signs; the $5d$ threshold test alone does not establish an eigenvalue. \\
Multicenter small-mass limit & Curvature term neglected as $\mu\to0$ & The corrected curvature contribution survives; the limiting Friedrichs operator has no normalizable zero mode. \\
Molecular separation & Analytical separated far-field solutions & The formal truncation omits terms of the retained asymptotic order and does not determine a molecular spectrum. \\
\hline\hline
\end{tabular}\endgroup
\end{table*}
\section{Conclusion}
Table~\ref{tab:corrections} summarizes the equation-level corrections, while Table~\ref{tab:physics} distinguishes their spectral and physical implications from the algebraic inconsistencies themselves.
Within the spatial model of Ref.~\cite{ISS}, the negative curvature sign drives the central inverse-square interaction beyond the Hardy threshold: no self-adjoint realization reaching the center has a ground state, whereas the positive sign admits a lower-bounded Friedrichs extension. At extremality the additional throat continuum excludes the reported $3p$ energy as a normalizable exterior bound state. Coordinate density peaks alone do not demonstrate trapping. In multicenter geometries, the curvature term persists in the formal small-mass limit, and the corrected separated Legendre solutions do not establish a molecular spectrum. Quantifying the impact on the published numerical profiles requires solving the corrected spectral problems with specified operator domains.

\section*{AI Use Disclosure}
OpenAI's ChatGPT (GPT-6.2) was used to assist with language editing, phrasing refinement, and \LaTeX{} formatting during manuscript preparation. The author directed and verified all edits and takes full responsibility for the scientific content and conclusions of this work.

\end{document}